\documentclass[a4paper,10pt,twocolumn,preprint,3p]{elsarticle}
\usepackage{lineno,hyperref}
\modulolinenumbers[5]

\journal{Journal of Electron Spectroscopy and Related Phenomena}

\biboptions{sort&compress}
\usepackage{amsmath}
\usepackage{amsfonts}
\usepackage{color}
\usepackage{ulem}

\def\bra#1{\mathinner{\langle{#1}|}}
\def\ket#1{\mathinner{|{#1}\rangle}}
\newcommand{\braket}[2]{\langle #1|#2\rangle}
\newcommand{\ketbra}[2]{|#1\rangle\langle#2|}

\def\sgn{\mathrm{sgn}}
\def\vv#1{\mathbf{#1}}

\def\GS{\text{GS}}
\def\creation#1{\hat{c}_{#1}^{\dagger}}
\def\annihilation#1{\hat{c}_{#1}}
\def\fieldCreation#1{\hat c_{#1}^\dagger}
\def\fieldAnnihilation#1{\hat c_{#1}}
\def\vacuum{\ket{0}}

\def\ketgs{| \Psi_0^{N} \rangle}
\def\brags{\bra{\Psi_0^{N}}}
\def\brap{\langle \Psi_n^{N+1}|}
\def\bram{\langle \Psi_n^{N-1}|}
\def\ketp{| \Psi_n^{N+1} \rangle}
\def\ketm{| \Psi_n^{N-1} \rangle}
\def\ketnn{| \Psi_n^{N} \rangle}

\begin{document}

\begin{frontmatter}

\title{Single- and many-particle description of scanning tunneling spectroscopy}

\author[mymainaddress]{Mikko M. Ervasti}
\author[nanoaddress]{Fabian Schulz}
\author[nanoaddress]{Peter Liljeroth}
\author[mymainaddress]{Ari Harju\corref{cor1}}
\ead{ari.harju@aalto.fi}
\cortext[cor1]{Corresponding author}
\address[mymainaddress]{COMP Centre of Excellence, Department of Applied Physics, \\Aalto University School of Science, PO Box 11100, 00076 Aalto, Finland }
\address[nanoaddress]{Department of Applied Physics, Aalto University School of Science, PO Box 15100, 00076 Aalto, Finland}

\begin{abstract}
Scanning tunneling spectroscopy measures how a single electron with definite energy propagates between a sample surface and the tip of a scanning tunneling microscope. In the simplest description, the differential conductance measured is interpreted as the local density of states of the sample at the tip position. This picture, however, is insufficient in some cases, since especially smaller molecules weakly coupled with the substrate tend to have strong Coulomb interactions when an electron is inserted or removed at the molecule. We present theoretical approaches to go from the non-interacting and single-particle picture to the correlated many-body regime. The methodology is used to understand recent experiments on finite armchair graphene nanoribbons and phthalocyanines. We also theoretically discuss the strongly-correlated model system of fractional quantum Hall droplets.
\end{abstract}


\end{frontmatter}


\section{Introduction}

Scanning tunneling spectroscopy (STS) combines atomic spatial resolution with current-voltage spectroscopy, making it possible to probe the electronic structure of single molecules and atoms
\cite{Binnig1982_HelPhysAc, Chen2008, Lyo1989_Science, Stipe1998_Science, Krull2013_NMat, Nilius2002_Science, Serrate2010_NNano, Ruby2015_PRL}.
Such experiments are of fundamental importance for understanding charge and spin transport in nanoscale systems. STS is based on the scanning tunneling microscope {(STM)}, an instrument that measures the quantum-mechanical tunneling current between a conducting sample and a sharp probe tip to generate a topographic image of the sample surface. As will be discussed in detail in this review, the tunneling current depends on the overlap between the tip and sample wave functions, and measurement of the tunneling conductance as a function of applied bias voltage allows deducing the local density of states (LDOS) of the sample.

Measurements on molecular systems are usually interpreted in terms of single-particle molecular orbitals \cite{PhysRevLett.94.026803, Liljeroth2010, Albrecht2013}.
{However, while optical spectroscopy probes the excitation spectrum of a molecule with a fixed number of electrons, STS measures the spectrum in case an electron is inserted or removed, due to tunneling between the tip and molecule.
If the molecule is small enough, the charging energy to overcome the Coulomb repulsion of the electrons of the neutral molecule can be substantial.
Therefore, to describe the eigenstates and -energies of especially the charged molecule properly, it is not clear \textit{a priori} whether the single-particle interpretation suffices.}
Even so, many-body effects, which have been studied in detail in correlated electron systems such as (high Tc) superconductors \cite{Hanaguri2004, Gomes2007}, are usually neglected.
{The more general quantity that is applicable beyond the single-particle description, and one that corresponds to LDOS and the STS $dI/dV$ measurements, is the single-particle spectral function \cite{bruus}. It will be discussed extensively in the current review.}

{In order to compare to the experimental measurements, one can do theory and calculations in various levels of sophistication. We will discuss mainly the three cases, namely the non-interacting, the single-particle (or mean-field or independent particles), and the interacting many-body picture. The various pictures can be defined by stating that in the first two the states are written as single Slater determinants, namely assuring that the particles occupy well defined molecular orbitals, whereas in the interacting case the states can be strongly correlated. In the non-interacting picture, the energies can be written as sums of occupied molecular orbital energies, whereas in the other two pictures the energies can be evaluated in principle using the full interacting Hamiltonian. We compare the theory that results in various pictures and present examples of systems where a more sophisticated description is needed. In the experimental side, it is interesting to find molecular systems that have visible features that can be explained only in the many-body picture. In the theory side, it is important to understand when a proper treatment of correlations and methods that go beyond the mean-field approaches, such as the density-functional theory (DFT), are required.}

{Furthermore, STS can potentially measure also higher-order tunneling processes between the tip, molecule and substrate, which can result in coherent many-body transport phenomena, such as cotunneling and the Kondo effect. For instance,}
in the Kondo effect a many-body singlet ground state forms between an unpaired spin in a molecule and a screening cloud of quasi-free electrons \cite{Zhao2005, Fernandez2008}.
{However, such processes go beyond the description of the states on the molecular sample alone, and they are out of scope in the current review, as we will solely} focus on observing intrinsic many-body effects on molecular systems using scanning tunneling spectroscopy.

{This Review is organized as follows. Section 2 gives a soft introduction to scanning tunneling spectroscopy. The spectral function is discussed in various levels of theory in Section 3. This consists of the non-interacting, single-particle, and interacting descriptions of the sample. Section 4 discusses the effect of the substrate on the STS measurements, comparing especially the cases of a thin insulating substrate layer to a fully metallic substrate. Metallic substrates tend to bring the molecule ionization (HOMO) and affinity (LUMO) levels closer together. Section 5 introduces example systems of finite armchair graphene nanoribbons, phthalocyanines, and strongly-correlated fractional quantum Hall droplets on various substrates. Finally, Section 6 concludes.}

\section{Basic principles of scanning tunneling spectroscopy}

{STM} is based on measuring the tunneling current flowing between an atomically sharp probe ("tip") and a conductive sample. In the simple approximation of a metal-vacuum-metal tunneling junction, and assuming the work function $\phi$ of tip and sample are equal, the current $I$ across the barrier is given by
\begin{equation}
I \propto e^{-2\kappa z} \quad \mathrm{with}\quad \kappa=\sqrt{2m\phi/\hbar} \ ,
\label{eq:decay}
\end{equation}
where {$z$ is the tip-sample distance}, $\hbar$ is Planck's constant divided by $2\pi$, and $m$ the mass of the electron. For typical values of metallic work functions of around 5 eV, the current will decrease by one order of magnitude if the tip-sample distance is increased by 1 \AA. Due to this exponential decay, essentially all of the electrons tunnel between the last atom of the tip and the substrate. This is the reason why STM can be used to obtain atomically resolved images of the sample surface \cite{PhysRevLett.50.120,Wiesendanger1994,Chen2008}.

STM imaging is typically carried out in the constant-current mode, where a feedback loop is used to adjust the $z$-position of the tip in order to maintain a constant tunneling current. Recording the $z$-signal yields a topography image of the sample surface. While this is related to the actual topography of the sample, it is also influenced by electronic effects as discussed below. Alternatively, the experiment can be carried out in constant-height mode by keeping
the height of the tip constant and recording the changes in the tunneling current. This mode is useful on atomically flat samples and in the case of examining flat molecules on well-defined surfaces.

In addition to the tip-sample distance, the tunneling current depends on the local density of electronic states (LDOS) $\rho_S$ close to the Fermi level of the sample. The sample LDOS can be probed in great detail by tunneling spectroscopy, i.e. by measuring the current as a function of the bias voltage. Understanding tunneling spectroscopy usually starts with the Bardeen picture of the tunneling process \cite{PhysRevLett.6.57}, where the current {can be expressed in terms of the tunneling matrix element $M_{\mu\nu}$  between states $\psi_\mu$ in the tip and $\psi_\nu$ in the sample:}
\begin{align}
I = &\frac{2\pi e}{\hbar^2}\sum_{\mu,\nu} f(E_\mu)[1-f(E_\nu+eV)]  \nonumber \\
&\times \vert M_{\mu\nu}\vert^2\delta(E_\mu-E_\nu)
\label{eq:BardeenI}
\end{align}
where $f(E)$ is the Fermi function, $V$ the applied bias between the sample and the tip, and $E_\mu$ and $E_\nu$ are the energies of states $\mu$ and $\nu$. The tunneling matrix element is given by
\begin{equation}
M_{\mu\nu} = \frac{\hbar^2}{2m}\int_{S} (\psi_{\mu}^\ast \nabla \psi_\nu - \psi_\nu \nabla \psi_\mu^\ast ) \cdot\mathrm{d} \textbf{S} ,
\label{eq:BardeenM}
\end{equation}
where the surface $S$ can be any surface lying within the barrier. This theory can be adapted to modeling STM as shown by Tersoff and Hamann \cite{PhysRevLett.50.1998,PhysRevB.31.805}. They expressed  the surface wave function $\psi_{\nu}$ in a very general form as a sum over Bloch waves in the plane of the surface, decaying exponentially into the vacuum.
{The tip was treated} by approximating it as a locally spherical potential well while neglecting its detailed atomic structure.
In the barrier region, the tip wave function $\psi_{\mu}$ thus exhibits an asymptotic spherical form around the center of curvature of the tip $\mathbf{r}_0$. This yields as the matrix element
\begin{equation}
M_{\mu\nu} \propto \psi_{\nu}(\mathbf{r}_0) .
\end{equation}
{This is known as the s-wave tip model. This matrix element is modified if the higher angular momentum orbitals of the tip are important \cite{Chen2008}. For example, for a $p_x$-wave tip}
\begin{equation}
M_{\mu\nu} \propto \frac{\partial\psi_{\nu}(\mathbf{r}_0)}{\partial x}
\end{equation}
The current for finite bias voltage and temperature is the convolution of the tip and surface DOS integrated over energy (converting the sum in Eq. \ref{eq:BardeenI} into an integrals)\cite{Chen2008}
\begin{align}
I = \frac{4\pi\mathrm{e}}{\hbar} \int_{-\infty}^{\infty} \big[f(E_F - \mathrm{e}V + \epsilon) - f(E_F + \epsilon)\big] \nonumber \\
 \rho_s(\mathbf{r_0},E_F - \mathrm{e}V + \epsilon)\rho_t(E_F + \epsilon) T(\epsilon,V)\mathrm{d}\epsilon ,
\label{eq:I}
\end{align}
where $\rho_t(E)$ is the tip density of states, transmission $T(E,V)$ contains the energy dependence of the matrix elements and
\begin{equation} \label{eq:ldos}
\rho_s(\mathbf{r_0},E)=\sum_\nu\vert\psi_{\nu}(\textbf{r}_0)\vert^2\delta(E_{\nu}-E)
\end{equation}
is the local density of states (LDOS) probed at the position of the tip.
{At low temperatures,} and if $\rho_t$ and $T$ do not depend (strongly) on energy, the differential conductance d$I$/d$V$ is proportional to the sample DOS at voltage $V$:
\begin{equation}
\frac{\mathrm{d}I}{\mathrm{d}V} \propto \rho_s(E_F + \mathrm{e}V)
\label{eq:dIdV}
\end{equation}
This very important result forms the basis for all scanning tunneling spectroscopy (STS) experiments, where the d$I$/d$V$ signal is interpreted as the LDOS of the sample that can be probed with atomic resolution. While different procedures for recovering the LDOS rigorously from the measured d$I$/d$V$ have been discussed in the literature \cite{PhysRevB.75.235432,RevSciInstrum.79.043104,PhysRevB.80.165419,PhysRevB.79.045404}, the interpretation of  d$I$/d$V \propto \rho_s(E_F + \mathrm{e}V)$ is usually sufficient. In real life, the instruments have finite energy resolution (due to finite temperature, electrical noise and other broadening mechanisms) and the LDOS should be considered as a sum of the states in the sample over the energy resolution of the instrument $\delta E$: $\rho_s(E_F+\mathrm{e}V)\approx\sum_{\delta E}\vert\psi_{\nu}(\textbf{r}_0)\vert^2\delta(E_{\nu}-E)$.

\section{Spectral function in various levels of theory}

{
In the standard description of STS, the electrons are assumed to tunnel between the tip and sample one by one.
Therefore the proper quantity that simulates the $dI/dV$ measurements is, strictly speaking, not the single-particle LDOS, but the more general single-particle spectral function \cite{bruus}
}
\begin{equation} \label{eq:def_T0_spectral_function}
A(\nu, \omega) = A^{-}(\nu, \omega) + A^{+}(\nu, \omega) ,
\end{equation}
{where we have separated the negative (annihilation) and positive (creation) side of the spectra to their respective functions. At zero temperature, they can be written as}
\begin{align}
&A^{-}(\nu, \omega) \nonumber \\
&= \sum_{n}
	\Big| \bram \annihilation{\nu} \ketgs \Big|^2
	\delta(\omega - (E_{0}^N - E_{n}^{N-1})) , \label{eq:A_annihilation} \\
&A^{+}(\nu, \omega)  \nonumber \\
&= \sum_{n}
	\Big| \brap \creation{\nu} \ketgs \Big|^2
	\delta(\omega - (E_{n}^{N+1} - E_{0}^N )) , \label{eq:A_creation} 
\end{align}
{where $\ketnn$ is the $n$:th many-body eigenstate of the sample with $N$ particles, $E_{n}^{N}$ is the energy of the state $\ketnn$, $\omega$ is the energy that corresponds to the bias voltage in STM, and $\nu$ labels the single-particle orbital where an electron is annihilated or created. In practice, $\ket{\nu}$ often corresponds to a position eigenket $\ket{\vv{r}}$, where the tip is positioned.}

{The spectral function can be intuitively thought of as electrons being annihilated on the sample at negative sample bias $\omega$ and created at positive bias. Here, the spin is not explicitly visible in the equation, but one typically has to separately consider all cases, where either spin-up or spin-down electron is created or annihilated. The energy gap between the first creation (affinity) and annihilation (ionization) peaks (energy levels) is the fundamental gap $\Delta = E^{N+1}_{0} - 2 E^{N}_{0} + E^{N-1}_{0}$, which is also the second order central finite difference of the charging curve, namely the total energy of the system as a function of the excess charge.}

\subsection{Non-interacting particles: From spectral function to LDOS}

We will next get familiar with this formula and show first that Eq.~(\ref{eq:def_T0_spectral_function}) reduces to
{the LDOS defined in Eq.~(\ref{eq:ldos}) for non-interacting particles.}
Without interactions, $\ketgs$ is simply
{a totally} antisymmetric
state where $N$
lowest-energy single-particle states are occupied. In the second quantization notation, the $i$th single-particle state is
{$\ket{\phi_i}=\creation{i} \vacuum$} and 
{it has a}
coordinate space wave function 
{$\phi_i(\vv{r})=\braket{\vv{r}}{\phi_i}$.}
We assume this single-particle state to have energy $\varepsilon_i$.
The $N$-particle ground-state is then $\ketgs=\prod_{i=1}^{N}\creation{i} \vacuum$ and it has energy $E_{0}^N=\sum_{i=1}^N \varepsilon_i$.

Next we start from the non-interacting $N$-particle ground-state $\ketgs$, and we would like to obtain the rate for one of the particles to
tunnel out from the system.
The spectral density to tunnel out from a single-particle state $\ket{\nu}$ is given by
$$ \sum_{n}
\Big| \bram \annihilation{\nu} \ketgs \Big|^2
\delta(\omega - (E_{0}^N - E_{n}^{N-1}))  , $$
where state $\ketm $ has one less particle. The non-interacting eigenstates of $N-1$ particles are simply given by configurations
$\prod_{i=1}^{N-1}\creation{\eta_i} \vacuum$, where the occupied orbitals are given by a set of indices $\{\eta_i\}_{i=1}^{N-1}$,
different for each state $\ketm $.
To get a non-zero contribution for
$\Big| \bram \annihilation{\nu} \ketgs \Big|^2$, the configurations $\ketm$ and $\ketgs$ can only differ by one occupation.
This means that the occupations of $\ketm$ have to be $N-1$ indices between one and $N$
(and because of fermions, the occupations can not be the same for two particles)  and $\nu$ has to be the index of the remaining occupied state.
In other words, the $N$ configurations giving non-zero contribution are simply
$\annihilation{\nu} \ketgs$, with $1 \le \nu \le N$.
For each fixed $\nu$ that corresponds to an occupied state in $\ketgs$,
the overlap is one, {meaning} that all the tunneling peaks have the same height. The peak positions are given by
$\delta(\omega - (E_{0}^N - E_{\nu}^{N-1})) $. Because the particles are here non-interacting, the total energy is a sum of the occupied
single-particle state energies.
The energy of the state $\ketm$ is simple as it differs from $\ketgs$ by occupation
of single-particle state $\nu$ only, and thus the total energy is $E_{\GS}-\varepsilon_\nu$.
{Therefore}
$\delta(\omega - (E_{0}^N - E_{n}^{N-1}))  = \delta(\omega - \varepsilon_{\nu})$, where $1\le \nu \le N$.
The tunneling  is thus possible at energies that correspond to the $\nu$th occupied single-particle energy.

If we want to find the tunneling amplitude for a specific point in space, we can again use Eq.~(\ref{eq:def_T0_spectral_function}). One should then
consider $\nu$ to be a state in the position basis. The operator for annihilating a particle at certain position is the field operator, and the standard
notation would be $\hat \psi(\vv{r})$ and for creation $\hat \psi^\dagger(\vv{r})$ (see, e.g., \cite{9781139023979}), but $\hat c_{\vv{r}}$ and $\hat c_{\vv{r}}^\dagger$ could also be used.
One can expand the field operator in a single-particle basis as $\hat c_{\vv{r}}=\sum_i \phi_i(\vv{r}) \hat c_i$, where $\phi_i(\vv{r})$ is the $i$th single-particle state. Inserting this {into}  Eq.~(\ref{eq:def_T0_spectral_function}) and repeating the
same calculation as above, one gets that again the tunneling at energies corresponding to the eigenstates, but now the peak height corresponding
to the $i$th single-particle state is given by $| \phi_i (\vv{r})|^2$. This is exactly what the local density of states calculation would give:
\begin{equation}
A^{-}(\vv{r}, \omega)= \sum_{i=1}^N |\phi_i(\vv{r})|^2 \delta(\omega-\varepsilon_i) .
\end{equation}
Note that the summation in this formula is over the $N$ occupied single-particle states.

In a similar fashion, one can consider particles tunneling {into} a ground state of $N$ particles. The relevant part of spectral function is then
$$ \sum_{n}
	\Big| \brap \creation{\nu} \ketgs \Big|^2
	\delta(\omega - (E_{n}^{N+1} - E_{0}^N )) , $$
where the state $\ketp$ is a non-interacting eigenstate with $N+1$ particles.  The states in the sum that give non-zero contribution now are
$\ketp=\creation{N+n} \ketgs$, with $n > 0$, meaning that the added particle occupies an originally unoccupied orbital with index $N+n$.
The total energy is again the sum of the single-particle energies and given by $E_{n}=E_{\GS}+\varepsilon_{N+n}$. Tunneling is thus possible
for each unoccupied single-particle eigenstate with same amplitude and energy that corresponds to single-particle energies $\varepsilon_j$ with
$j>N$.

Using the results above and expansion $c_{\vv{r}}^\dagger=\sum_i \phi^*_i(\vv{r}) c_i^\dagger$, the spectral function for tunneling in
at position $\vv{r}$ can be shown to be
\begin{equation}
A^{+}(\vv{r}, \omega)= \sum_{i=N+1}^\infty |\phi_i(\vv{r})|^2 \delta(\omega-\varepsilon_i) .
\end{equation}
It is interesting to note that this equation complements the spectral function of tunneling out in a sense that when both effects are combined, one gets
tunneling at occupied and empty states in a single sum as
\begin{equation}
A(\vv{r}, \omega)= \sum_{i=1}^\infty |\phi_i(\vv{r})|^2 \delta(\omega-\varepsilon_i) .
\end{equation}
{This shows} that the spectral function leads to the same results as the local density of states for the non-interacting
particles.

\subsection{Mean-field description}

To go beyond a simple non-interacting particle picture, we will first assume that the correlation effects are small and a mean-field approximation is sufficient.
The peak positions in Eq.~(\ref{eq:def_T0_spectral_function}) come from the total energies of the $N$ and $N \pm 1$ particle systems,
and the Coulomb energy part of these is not captured by a simple single-particle description, but approximately included by the
mean-field.
In  doing a mean-field approximation for the many-body problem, one still considers the system to be described
{by a single Slater determinant, as}
 was done for the non-interacting ground states, but one now optimizes the occupied single-particle orbitals to minimize the total energy.
If we do separately the mean-field calculation for the $N$ and $N\pm 1$ particle systems, we obtain the total energies for these systems and also different single-particle orbitals in each case.
The total energy differences give the tunneling peak energies.
To calculate the matrix elements, it is first helpful to define the $N$-particle mean-field ground state to be
$\ketgs = \creation{\eta_1} \dots \creation{\eta_N} \ket{0}$ and for the
$N-1$ particle eigenstate $\creation{{\mu}_{1}} \cdots \creation{{\mu}_{N-1}} \ket{0} $.
Notice that this state does not have to be the ground state. With these definitions, the matrix elements are
\begin{align}
&\brags\creation{p} \ket{\Psi_{n}^{N-1}}\nonumber \\
=& \bra{0} \big[ \annihilation{\eta_N} \cdots \annihilation{\eta_1} \big]
\creation{p}
\big[ \creation{{\mu}_{1}} \cdots \creation{{\mu}_{N-1}} \big] \ket{0} \nonumber \\
=& \sum_{i_1, i_2, \hdots, i_N} \braket{\eta_{i_1}}{p}
\Big[ \braket{\eta_{i_2}}{\mu_1} \cdots \braket{\eta_{i_N}}{\mu_{N-1}} \Big] \nonumber\\
&\times \bra{0} \annihilation{\eta_N} \cdots \annihilation{\eta_1}
\creation{\eta_{i_1}} \creation{\eta_{{i}_{2}}} \cdots \creation{\eta_{{i}_{N}}} \ket{0} \nonumber \\
=& \sum_{P \in S_N} \sgn(P) \;
\braket{\eta_{P(1)}}{p} \nonumber \\
&\times \Big[ \braket{\eta_{P(2)}}{\mu_1} \cdots \braket{\eta_{P(N)}}{\mu_{N-1}} \Big] \nonumber \\
=&\det
\begin{pmatrix}
    \braket{\eta_1}{p} & \braket{\eta_1}{\mu_1} & \cdots & \braket{\eta_1}{\mu_{N-1}} \\
    \braket{\eta_2}{p} & \braket{\eta_2}{\mu_1} & \cdots & \braket{\eta_2}{\mu_{N-1}} \\ 
    \vdots        & \vdots  & \ddots & \vdots  \\ 
    \braket{\eta_N}{p} & \braket{\eta_N}{\mu_1} & \cdots & \braket{\eta_N}{\mu_{N-1}} \nonumber
\end{pmatrix} .
\end{align}
This means that one needs just to calculate the overlaps of the single-particle orbitals from the mean-field calculations.
The other term needed is $\brags\annihilation{p} \ket{\Psi_{n}^{N+1}} = \left[ \bra{\Psi_{n}^{N+1}} \creation{p} \ketgs \right]^\dagger$, which can be readily evaluated by using the formula above.

One can also make an approximation that after the mean-field calculation is performed at particle number $N$, the same single-particle
orbitals are used for the $N \pm 1$ states. This leads to a similar formula that was used for non-interacting particles, namely, 
{one can then}
calculate the LDOS from the mean-field orbitals. This is actually what most calculations based on DFT do.

\subsection{Spectral function in correlated systems}

{The spectral function defined in Eq.~(\ref{eq:def_T0_spectral_function}) can be used in the interacting many-body case that goes beyond the mean-field approximation.}
Any correlated many-body wave function can 
be written as a linear combination of the Slater determinants that are the $N$-particle configurations
$\ket{\Phi_i^N} = \Big( \prod_{j=1}^{N}\creation{i_j} \Big) \vacuum$, where 
$\ket{\phi_i}=\creation{i} \vacuum$ are the single-particle orbitals used in the expansion. Explicitly, an arbitrary many-body state is written as
$$
\ket{\Psi_n^N} = \sum_i \alpha_i^n \ket{\Phi_i^N} .
$$
The coefficients $\{ \alpha_i^n \}_{i=1}^M$ for the $n$th many-body eigenstate can be obtained by solving the many-body Schr\"odinger equation. The same solution also gives the many-body energies $E_{n}^{N}$.
The relative energies of the many-body eigenstates determine the peak positions in the spectrum.
On the other hand, the peak weights at a position $ \vv{r}$ correspond to the squared amplitudes of terms like
\begin{align}
& \bra{\Psi^{N-1}} \fieldAnnihilation{ \vv{r}} \ket{\Psi^{N}} \nonumber \\
=& \frac{1}{N!}
\int d \vv{r}_{1} \hdots d \vv{r}_{N}\nonumber \\
&\times
\bra{\Psi^{N-1}}
\fieldAnnihilation{ \vv{r}}
\ket{ \vv{r}_{1} \hdots  \vv{r}_{N}}
\braket{ \vv{r}_{1} \hdots  \vv{r}_{N}}{\Psi^{N}} \nonumber \\
=& \frac{1}{N!}
\int d \vv{r}_{1} \hdots d \vv{r}_{N}
\sum_{i} \delta( \vv{r} -  \vv{r}_{i}) \nonumber \\
&\times
\braket{\Psi^{N-1}}{ \vv{r}_{1} \hdots \underline{ \vv{r}_{i}} \hdots  \vv{r}_{N}}
\braket{ \vv{r}_{i},  \vv{r}_{1} \hdots \underline{ \vv{r}_{i}} \hdots  \vv{r}_{N}}{\Psi^{N}} \nonumber \\
=&
\frac{1}{(N-1)!}
\int d \vv{r}_{2} \hdots d \vv{r}_{N}\nonumber \\
&\times\Psi^{N-1}( \vv{r}_{2} \hdots  \vv{r}_{N})^{*}
\Psi^{N}( \vv{r},  \vv{r}_{2} \hdots  \vv{r}_{N}) , \label{eq:A_amplitude_derivation}
\end{align}
where $\underline{ \vv{r}_{i}}$ denotes a missing coordinate.
One should note that the prefactors in the formulas depend on the normalization of the coordinate representation many-body wave functions and we
have followed the convention of \cite{9781139023979}.
 Moreover, the terms with the creation operator can be evaluated similarly as $|\bra{\Psi^{N}} \fieldCreation{ \vv{r}} \ket{\Psi^{N-1}}|^2 = |\bra{\Psi^{N-1}} \fieldAnnihilation{ \vv{r}} \ket{\Psi^{N}}|^2$.

The spectral function is effectively a single particle quantity such that the terms $\bra{\Psi^{N-1}} \fieldAnnihilation{\vv{r}} \ket{\Psi^{N}}$ can be expanded using single-particle orbitals. For instance, expanding using the natural orbitals $\phi_{i}^{1}( \vv{r})$
{(the eigenfunctions of the one-body reduced density matrix)}, the natural $(N-1)$-states $\phi_{i}^{N-1}( \vv{r}_{2}, \hdots  \vv{r}_{N})$, and their common eigenvalues $|c_i|^2$ of the many-body ground state $\ket{\Psi^{N}}$, Eq.~(\ref{eq:A_amplitude_derivation}) can be written as
\cite{Coleman_1963}
\begin{equation} \label{eq:mb_amplitudes}
\bra{\Psi^{N-1}} \fieldAnnihilation{\vv{r}} \ket{\Psi^{N}}
= \sum_{i} c_{i} \braket{\Psi^{N-1}}{\phi_{i}^{N-1}} \phi_{i}^{1}( \vv{r}).
\end{equation}
The many-body effects on the spectral function are then more visible as the correlations of the many-body ground state that results in a wide distribution of the natural orbital occupations $|c_i|^2$ and the overlaps $\braket{\Psi^{N-1}}{\phi^{N-1}}$. If the natural orbitals are solved first, the natural $(N-1)$-states can be solved for instance using the identity
\begin{equation}
c_i \phi_{i}^{N-1}( \vv{r}_{2}, \hdots ,  \vv{r}_{N}) = \int d \vv{r} \: \phi_{i}^{1}( \vv{r})^{*} \Psi( \vv{r},  \vv{r}_{2}, \hdots,  \vv{r}_{N}) . \nonumber
\end{equation}
{Furthermore, as shown above, the spectral function can be formally written as
$$A(\nu,\omega) = \bra{\vv{r}} \sum_i \hat{\rho}^{1}_{i} \delta(\omega - E_i) \ket{\vv{r}}, $$
where $\hat{\rho}^{1}_{i} = \sum_{ij} d_i^* d_j \ketbra{\phi^{1}_i}{\phi^{1}_j}$ and the coefficients $d_i$ are defined by Eq.~(\ref{eq:mb_amplitudes}), and $E_i$ are the peak positions that depend on $E^{N}_0$, $E^{N+1}_n$ and $E^{N-1}_n$.
That is, the spectral function can be interpreted as a single-particle density matrix that is distributed in energy. In the independent particle picture the LDOS is formally of the same form, where $\hat{\rho}^{1}$ correspond to pure states. The mixed state of the propagating electrons reflect the correlated many-body properties of the molecular states.}

\subsubsection{Monte Carlo integration}

For the cases where the many-body wave function has an analytic form,
the integral in Eq.~(\ref{eq:A_amplitude_derivation}) can still be tedious to calculate for large particle numbers. Monte Carlo integration with importance
sampling is one obvious option for this integral. This is based on the fact that
\begin{align}
&\int d \vv{r}_{2} \hdots d \vv{r}_{N}\nonumber 
\Psi^{N-1}( \vv{r}_{2} \hdots  \vv{r}_{N})^{*}
\Psi^{N}( \vv{r},  \vv{r}_{2} \hdots  \vv{r}_{N}) \nonumber \\
&/ \int d \vv{r}_{2} \hdots d \vv{r}_{N} |\Psi^{N-1}( \vv{r}_{2} \hdots  \vv{r}_{N})|^2 = \nonumber \\
=&\int d \vv{r}_{2} \hdots d \vv{r}_{N}\nonumber 
|\Psi^{N-1}( \vv{r}_{2} \hdots  \vv{r}_{N})|^2
\frac{\Psi^{N}( \vv{r},  \vv{r}_{2} \hdots  \vv{r}_{N})}{\Psi^{N-1}( \vv{r}_{2} \hdots  \vv{r}_{N})} \nonumber \\
&/ \int d \vv{r}_{2} \hdots d \vv{r}_{N} |\Psi^{N-1}( \vv{r}_{2} \hdots  \vv{r}_{N})|^2  \nonumber \\
= & \left \langle \frac{\Psi^{N}( \vv{r},  \vv{r}_{2} \hdots  \vv{r}_{N})}{\Psi^{N-1}( \vv{r}_{2} \hdots  \vv{r}_{N})}\right \rangle_{|\Psi^{N-1}( \vv{r}_{2} \hdots  \vv{r}_{N})|^2} \ ,
\label{eq:MC}
\end{align}
where the last equation means that the coordinates $\vv{r}_{2} \hdots  \vv{r}_N$ are sampled from distribution
$|\Psi^{N-1}( \vv{r}_{2} \hdots  \vv{r}_{N})|^2$ and the remaining coordinate $\mathbf{r} $ specifies the point where the spectral function is evaluated at.
The wave function ratio in Eq.~(\ref{eq:MC}) should be straightforward to calculate for analytic functions.

\section{Substrate effects}

{
If the substrate underneath couples to the sample, it can induce a noticeable effect on the sample states and energies. In such a case the full system with both the sample and the substrate has to be considered. It is possible to introduce a thin insulating layer to decouple the molecular sample from the metallic substrate. We discuss both cases in what follows.
}

\subsection{{Metallic substrate}}

When a molecule is brought close to a metallic substrate, its energy levels can be drastically modified. Most notably, the energy gap between the ionization (HOMO) and affinity levels (LUMO) is reduced from that of an isolated molecule. This is due to:
\begin{enumerate}
\item One-body couplings between the molecule and substrate states shifting and broadening the energy levels, such as in the Fano model \cite{9781139023979}.
\item The electron-electron interactions between the molecule and the substrate perturbing [or induces changes in] the densities, which alters the molecule energy levels.
\item Electron or hole additions [or electron removals] on the molecule causing the electron states at the metallic substrate to relax further to minimize the total energy, resulting in charge polarization (dynamical screening). The correlation energies are heavily altered, and the gap between the ionization (HOMO) and affinity levels (LUMO) substantially decreases \cite{Neaton_2006}. Higher DOS at the Fermi level will results in even smaller energy gaps between the molecular energy levels \cite{Thygesen_2009}.
\item The molecule can become polarized {as well }due to dynamic charge transfer from a strongly-coupled metallic substrate, especially as a result of the electron or hole additions \cite{Thygesen_2009}.
\end{enumerate}
Even with an insulating layer between the molecule and the metallic substrate, the dynamical screening by the substrate can clearly reduce the energy gaps \cite{giant}. The polarization effects can be estimated by evaluating the screened Coulomb interactions or using classical image charge corrections \cite{Neaton_2006, Ruffieux20126930}. It should be noted that the Kohn-Sham or Hartree-Fock energy levels do not take the dynamical screening of added electrons or holes into account, and such single-particle picture is insufficient in general to describe the energy peak positions of molecules on a metallic substrates.

\subsection{{Insulating and weakly-coupled substrates}}

If the molecules are electronically sufficiently decoupled from the substrate (through e.g. the use of an ultrathin insulating film), the tip-molecule-substrate system can be modeled as a double-barrier tunnel junction
\cite{Science299.542,PhysRevLett.93.236802,PhysRevLett.94.026803,Nazin21062005,PhysRevB.93.121406,PhysRevB.85.205408,PhysRevB.86.155451}.
In this regime, the electronic states of the molecule are not hybridized with the metallic substrate and the charge in the molecule is quantized. The tunneling
is similar to single-charge tunneling in nanostructures, see e.g. \cite{grabert2013single,C4CS00204K}.
The resonances in the tunneling spectra correspond to (transient) addition/removal of electrons from/to the molecule. 
Importantly, now as the charge in the molecule is quantized, the total energies in Eq.~(\ref{eq:def_T0_spectral_function})
contain charging energies of the molecule.
In the gas phase of the molecule, these are the ionization potential (IP) and electron affinity (EA).
Of course, in the case of tunneling spectroscopy, the presence of the substrate has a significant effect on the electron-electron screening and additionally, depending on the relative alignment of the molecular resonances with the substrate Fermi level, it is also possible to observe permanent charging of the molecules
{\cite{PhysRevLett.106.216103, Steurer2015}}.

In the simple single-particle (constant interaction) picture, electrons will be added (at positive substrate bias) to the molecule when the bias voltage reaches the following condition (the tip Fermi level aligns with the lowest unoccupied molecular orbital, LUMO)
\begin{equation}\label{eq:LUMO}
\eta V=E_\mathrm{LUMO}-E_F+\Sigma_-
\end{equation}
where $\eta$ is the fraction of the applied bias that drops between the tip and the molecule and $\Sigma_-$ is the polarization energy associated with electron addition into the molecule. Correspondingly, at sufficiently negative bias, hole tunneling through the highest occupied molecular orbital HOMO can occur
\begin{equation}\label{eq:HOMO}
\eta V=E_\mathrm{HOMO}-E_F-\Sigma_+
\end{equation}
where $\Sigma_+$ is the polarization energy associated with positive charging of the molecule (sign taken as positive). These processes correspond to an opening of a new tunneling channel and result in a step increase in the measured tunneling current, i.e. a peak in the d$I$/d$V$ signal. The bias voltages corresponding to the resonances can be used to obtain the energies of the molecular orbitals w.r.t. the subtrate Fermi level. However, as the resonances in d$I$/d$V$ spectra correspond to temporary charging of the molecule, their energies are {affected} by the Coulomb energy involved in adding or removing one electron. This also implies that the STM transport gap $\Delta V_\mathrm{HOMO-LUMO}$ is not equal to the single-particle HOMO-LUMO gap (or the optical gap). They are related as follows:
\begin{align}
\label{eq:gaps}
\eta\Delta &V_\mathrm{HOMO-LUMO}= \nonumber \\
&E_\mathrm{LUMO}-E_\mathrm{HOMO}+\Sigma_-+\Sigma_+
\end{align}
{Again, the factor $\eta$ in Eq.~(\ref{eq:gaps}) results from the potential distribution in the double-barrier tunnel
junction and the finite bias drop between the molecule and the underlying metal substrate.}
Typically, the dielectric constant of the insulating film is large compared to vacuum and consequently, $\eta$ is close to unity,
{or often roughly $0.9$.}
This means that the bias voltage scale is almost equal to the real energy scale.

Finally, the STM transport gap also differs from the fundamental gap EA-IP due to the additional screening from the substrate, which affects the terms $\Sigma_+$ and $\Sigma_-$. If the molecules are further decoupled from the substrate through thicker insulating films, the STM gap approaches the fundamental gap \cite{PhysRevLett.94.026803}.

\section{Example systems}

In this section, the theory outlined above is used to model experiments on three example systems: finite graphene ribbons, phthalocyanine molecules, and
fractional quantum Hall droplets.

\subsection{{Finite armchair graphene nanoribbons}}

\begin{figure}
  \centering
  \includegraphics[width=\linewidth]{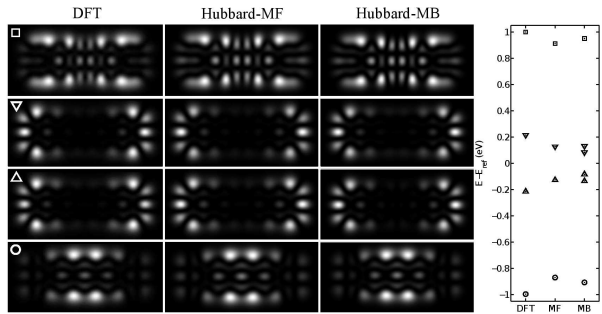}
  \caption{Simulated dI/dV maps of a three-monomer 7-AGNR from first principles (DFT), mean-field Hubbard model (Hubbard-MF), and the spectral function from the many-body Hubbard model (Hubbard-MB). The real-space imaging is simulated at a constant height of $3.5$\AA (DFT) and $4.0$\AA in (Hubbard-MF and Hubbard-MB). To produce the maps of the Hubbard models, each site is assumed to correspond to analytic $p_z$ orbitals with overlaps neglected. On the right, the corresponding STS spectra is shown. The many-body Hubbard model energy values are the ground state spectral function peak energies. From \cite{PhysRevB.88.075429}.
  }
  \label{fig:same}
\end{figure}

For carbon-based materials, one would not expect strong many-body phenomena, but still
the LDOS from a mean-field or density-functional-theory calculation can qualitatively differ from a many-body calculation. The main difference is that the spectral function peak energies come from the eigenenergies of the $N \pm 1$-particle systems in addition to the unperturbed $N$-particle system.
In Fig.~\ref{fig:same}, the simulated STM maps and spectra of a three-monomer-long and seven-atom-wide finite armchair graphene nanoribbon (7-AGNR) {are} shown. The LDOS maps from DFT and mean-field Hubbard model calculations agree well with the spectral function maps from an exact diagonalization calculation of the many-body Hubbard model. The agreement between the different computational strategies is surprisingly good.
However, the spectra do not fully match, since the spectral function has a double peak structure at both sides of the Fermi level ($E=0$). This discrepancy originates from the non-degenerate eigenenergies of the $N \pm 1$-particle systems that differ from the single-particle orbital energies of a neutral ribbon. In fact, the same double peak structure is present in the mean-field orbital energies in case of $N\pm 1$ particles. However, this is not shown in Fig.~\ref{fig:same} as the mean-field peak positions are not calculated from the total energies but from the mean-field orbital energies of the neutral 
system \cite{PhysRevB.88.075429}.
The conclusion of this is that the  correlation effects are too small to be resolved by present STM experiments, similarly
to Ref.~\cite{toroz} for the polyaromatic molecules.

\begin{figure}
  \centering
  \includegraphics[width=\linewidth]{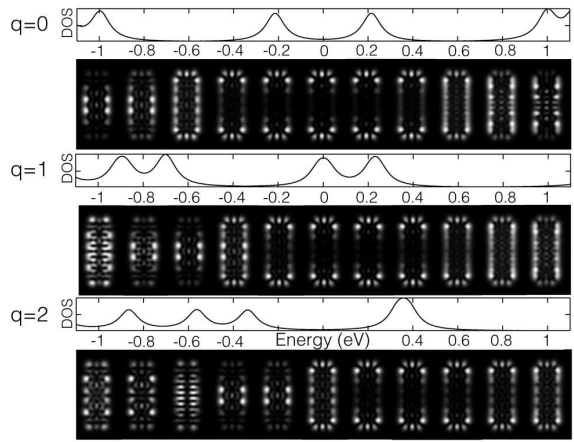}
  \caption{
  The DFT-calculated density of states (DOS) and simulated  dI/dV  maps for fully hydrogenated ribbons at different values of doping. The energy scale of the DOS figure applies also to the  dI/dV  maps in the sense that they range from  -1 to 1 eV in steps of 0.2 eV. 
  The q=0 is the antiferromagnetic uncharged system, q>0 are hole-doped systems with one and two holes, correspondingly. The density of states plots show the energies of the molecular orbitals, as well as indicate the overall magnitude of the individual  dI/dV maps (the maps are individually
  normalized which artificially enhances the signal between the peaks compared to experiments). From \cite{PhysRevB.88.075429} 
  }
  \label{fig:kukka}
\end{figure}

The effect of hole-doping on the finite 7-AGNRs is shown in Fig. \ref{fig:kukka}.
The neutral ($q=0$) AGNRs support antiferromagnetic (AFM) or ferromagnetic (FM) ordering between the zigzag-end-localized edges, the AFM order being only slightly lower in energy.
Doping the ribbons by removing one ($q=1$) or two ($q=2$) electrons quenches the magnetic order, that is resulting in spin-degenerate solution.
Moreover, the molecular orbital energies are altered, as the highest occupied orbitals become unoccupied by crossing to the unoccupied side above the Fermi level.
The energy gaps are also affected.
Especially the energy gap between the end-localized states (HOMO-LUMO gap in the $q=0$ case) becomes smaller, which evidently shows as a single broadened peak in the $q=2$ case at around $E=0.35$ eV.
The maps, however, show that the state ordering does not change by doping, and one can more or less understand the system by considering the single-particle picture with a changing Fermi level.
Experimentally, several different types of atomically precise graphene nanoribbons have been synthesized  \cite{ADMA:ADMA201505738}. When 7-AGNRs are deposited on Au(111) substrate, the maps with $q=2$ presented in Fig.~\ref{fig:kukka} provide the best match with the experiments \cite{Koch2012713,NatComm.4.2023}. The measurements show a single peak on the unoccupied side of the spectrum close to the Fermi level. The positive doping of the ribbons is consistent with the p-doping of graphene on a Au(111) substrate \cite{nn500396c}.

\begin{figure}
  \centering
  \includegraphics[width=\linewidth]{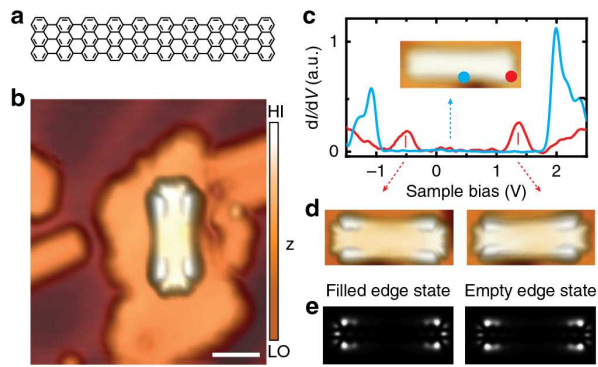}
  \caption{
  (a) Structural model of a (7, 20) GNR. Scale bar, 2 nm. (b) STM topography image of a (7, 20) GNR transferred onto a NaCl monolayer island through STM manipulation (U=-1.0 V, I=30 pA). Colour bar: HI=high; LO=low. (c) Differential conductance spectra measured in the centre (blue) and at a zigzag end (red) of the decoupled (7, 20) GNR. Inset: STM topography image at sample bias in the band gap of the ribbon (U=-0.5 V, I=30 pA). (d) STM topography images showing the orbital shapes of the occupied edge state (left, U=-1.0 V, I=30 pA) and the unoccupied edge state (right, U=1.4 V, I=30 pA). (e) Local density of states of corresponding Kohn-Sham orbitals at 4 \AA~distance above the GNR. From~\cite{giant}.
  }
  \label{fig:gnr_at_nacl}
\end{figure}

When the molecule is strongly coupled with a metallic substrate, the charging energy is expected to be small due to the strong screening. In addition, the charging energy depends on system size and can be negligibly small for large molecules such as graphene nanoribbons 
\cite{NatComm.4.2023,NatComm.6.10177}. Also, if the charging energy is small compared to the coupling with the substrate, the charge on the molecule is not quantized and it is possible to have molecular orbitals with non-integer occupation.
The nanoribbon gaps have been analyzed in detail in Ref.~\cite{Ruffieux20126930}.
If the molecules are further decoupled from the substrate through thicker insulating films, the STM gap approaches the fundamental gap 
\cite{PhysRevLett.94.026803}. {Such an experiment on graphene nanoribbons deposited on NaCl } is shown in Fig.~\ref{fig:gnr_at_nacl}. {In this case, the end states are prominently split, as expected for a neutral ribbon decoupled from the substrate. }
This measurement is in a nice agreement with the theoretical maps in Fig.~\ref{fig:same}. The energy resolution of the experiment is not sufficient to conclusively demonstrate the additional splitting of the end states predicted by the many-body calculation.

\subsection{Phthalocyanine}

\begin{figure}[!t]
  \includegraphics[width=\linewidth]{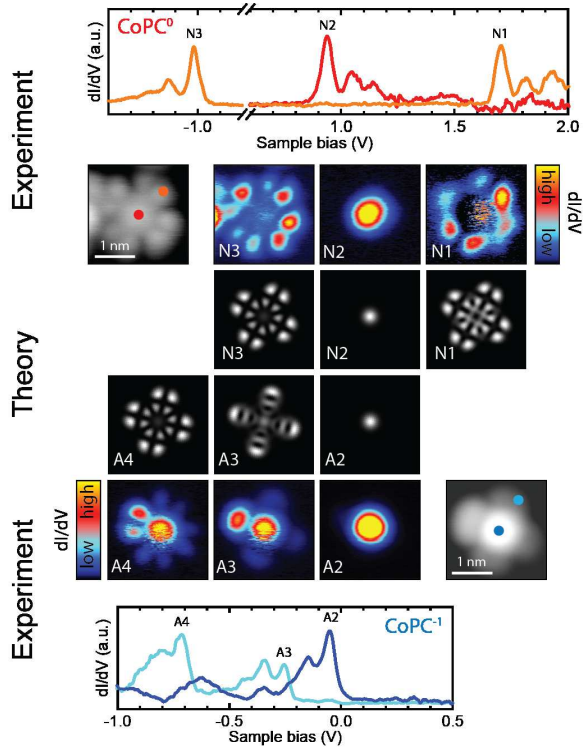}
  \caption{Comparison between experimental d$I$/d$V$ spectra (top-most and bottom-most row) and maps (second and fifth row) for neutral and negatively charged CoPC with excited states computed from time-dependent density functional theory (third and fourth row) \cite{Schulz2015229}}
  \label{fig:kops}
\end{figure}

{Molecules containing heavier atoms are expected to give rise to stronger correlation effects and we will consider metal phthalocyanine molecules as an example.
{It has been theoretically suggested that a many-body effect can be used to tune a low-spin to
high-spin transition in copper phthalocyanine (CuPc) deposited on an ultrathin insulating film under nonequilibrium conditions.
\cite{PhysRevB.93.121406,PhysRevB.85.205408,PhysRevB.86.155451}}}


Experimental results on cobalt phthalocyanine (CoPC) deposited on an hexagonal boron nitride monolayer (hBN) on Ir(111) are shown in Fig.~\ref{fig:kops} \cite{Schulz2015229}. Due to the work function modulation over the different sites of the hBN moir\'e pattern \cite{Schulz2014_PRB}, there are two different charge states of CoPC present on the surface \cite{Schulz201511121}. STS spectra for electron removal from (negative bias) and electron injection into (positive bias) neutral CoPC (CoPC$^{0}$) are shown in the up-most row in Fig. \ref{fig:kops}, while the bottom-most row shows spectra for electron removal from negatively charged CoPC (CoPC$^{-1}$). One would expect that CoPC$^{-1}$ would simply correspond to CoPC$^{0}$ through a downward shift in energy of all the observed transitions. This seems to hold true for peaks labelled A1 (at $\sim$0.8 V \cite{Schulz2015229}, not shown here) and A2 of the anionic molecule, which could be interpreted as peaks N1 and N2 of the neutral species shifted down by $\sim$1 eV. However, transition A3 of CoPC$^{-1}$, which would be expected to correspond to N3 of CoPC$^0$, should then appear well below -1.0 V. Instead, there are two transitions A3 and A4 just below A2 at energies of $\sim$-0.3 and $\sim$-0.8 V, respectively.
Note that all the transitions show satellite peaks due to electron-vibration coupling \cite{PhysRevLett.92.206102, Schulz201511121, PhysRevB.40.11834, Repp2010, NatComm.4.2023}.
In addition, maps of the spatial variation of the d$I$/d$V$ signal shown in the second and fifth row of Fig. \ref{fig:kops} reveal that the wave functions corresponding to transitions N3 and A3 exhibit a different shape and symmetry.

While resonances N2, N3 and A2 represent transitions between ground states of differently charged CoPC, A3 and A4 correspond to transitions into N-electron excited states (electron removal from CoPC$^{-1}$), which in the single-particle picture would be obtained by simply unoccupying the single-particle states of the molecule's ground state in their respective order, e.g. as given by a density functional theory calculation. However, this neglects the interaction of the hole with the remaining electrons, which can result in orbital re-ordering and mixing of multiple single-particle excitations. Such effects can be taken into account by using time-dependent density functional theory (TDDFT) \cite{Runge1984_PRL} to explicitly calculate the many-body excited states of the molecule and project them onto a chosen set of single-particle orbitals \cite{Casida2009_JMS}. The central rows in Fig. \ref{fig:kops} compare the dominant contributions to the TDDFT excited states upon projection onto the Kohn-Sham eigenstates of neutral CoPC with the experimental d$I$/d$V$ maps, and transitions A3 and A4 can be clearly identified with the first and second many-body excited states of neutral CoPC. The experimentally observed orbital re-ordering is well-described by the excited states computed from TDDFT, showing that this approach allows for a detailed understanding of STS spectra beyond the prevalent single-particle picture.

\subsection{Probing a fundamentally correlated state, example of $\nu=1/3$ Laughlin state}

\begin{figure}
  \centering
  \includegraphics[width=\linewidth]{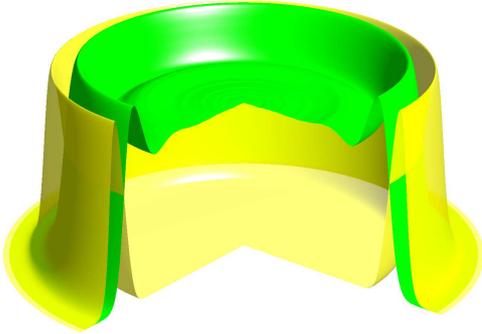}
  \caption{The particle density (green) and the spectral function (yellow) to add one particle to the system.
  One quarter has been removed to reveal the profiles more
  clearly. To reach a ground state with one more electron, the electron can tunnel only to a single orbital that actually corresponds to a non-interacting
  eigenstate of the model used.}
  \label{fig:laugh}
\end{figure}

We will end this section with a one of the best-known examples of strongly correlated state of matter, fractional quantum Hall droplet\cite{Ezawa}.
The Laughlin wave function for the $\nu=1/3$ state is given by \cite{PhysRevLett.50.1395}
\begin{align}
\Psi_L(\vv{r}_{1} \dots  \vv{r}_{N})&= \nonumber \\
&\prod_{i<j}^N (z_i - z_j)^3 \prod_{i=1}^N \exp(-r_i^2/4)  \ ,
\nonumber 
\end{align}
where $z=x + \mathrm{i} y$ and $r^2=x^2 + y^2$. Note that natural units for the problem are used.
Now Eq.~(\ref{eq:MC}) can be used to calculate the spectral function for the states with Laughlin wave function.
Fig.~\ref{fig:laugh} shows schematically the total particle density for 30 electrons and the spectral function for creating one extra particle, changing the system
to the ground state with one added electron. One can see that the bulk of the Laughlin state has a flat density but the edge has some structure. Further, the spectral function shows a structure that corresponds to a one single state in the lowest Landau level (for the function in the symmetric
gauge, see e.g. \cite{Ezawa}). One would expect that as each single-particle state in the system is $1/3$ occupied on average, the tunneling would have contribution from three single-particle orbitals. This is not the case, and the spectral function is identically zero for all orbitals except one.
One simple way
to understand this is based on the rotational symmetry and the resulting fact that the angular momentum of the quantum Hall droplet is a good
quantum number. Thus, the tunneling electron has to have angular momentum equal to the difference of the angular momenta
of the $N$ and $N+1$ particle systems. In the lowest Landau levels the orbitals are uniquely determined by the angular momentum, and
thus tunneling can only happen to one single-particle orbital.
{In conclusion, even in such a strongly-correlated model system such as the FQH droplet, the STS is not expected to see 
any correlated many-body features.}
Experimentally, STS has been used to study Landau levels and edge states in graphene, see \cite{Li}.

\section{Conclusions}

To conclude, we have shown how the theoretical modeling of scanning tunneling spectroscopy can be developed from the simple non-interacting particle picture to mean-field and full many-body description of the problem. 
For sample systems strongly contacted with a metallic substrate, the substrate can hinder the intrinsic many-body effects of the sample. Decoupling the
system from the metal by an insulating layer reduces the screening effects of the substrate.
Further, as the tunneling to metal is reduced, the system approaches a double-barrier setup and charge on the sample is closer to quantization, showing
charging effects. These charging effects can lead to level reorganization, perhaps the easiest many-body effect to see on scanning
tunneling spectroscopy.
We have used the theoretical tools 
to model three classes of example systems, namely graphene ribbons, phthalocyanines, and quantum Hall droplets. 
On graphene, many-body effects has been shown to be mostly missing, but phthalocyanines showed level reorganization. On the other hand, results
on quantum Hall droplets demonstrated the difficulty to capture many-body effects of this strongly correlated system, partially because scanning
tunneling spectroscopy at the level treated in this review is a single-particle probe.

\section*{Acknowledgements}

This research made use of the Aalto Nanomicroscopy Center (Aalto NMC) facilities and was supported by the European Research Council (ERC-2011-StG No. 278698 PRECISE-NANO), the Academy of Finland through its Centres of Excellence Program (projects no. 250280 and 251748), and the Finnish Academy of Science and Letters. We acknowledge the computational resources provided by Aalto Science-IT project and Finland's IT Center for Science (CSC).

\section*{References}

\bibliography{many_final}

\end{document}